\documentstyle[graphicx,prl,aps]{revtex}
   \newcommand{\cent}[1] {\begin{center}#1\end{center}}
   \newcommand{\doublint}{\int\rule{-3.5mm}{0mm}\int}
   \newcommand{\mra}  {\to}
   \newcommand{\vecbm}[1]{\mbox{\boldmath#1}}
   
   \newcommand{\goo}{\,\raisebox{-.5ex}{$\stackrel{>}{\scriptstyle\sim}$}\,}
   \newcommand{\lra}  {$\leftrightarrow$}
\begin{document}
\twocolumn
\draft
\title{
Phase transitions in finite systems = topological peculiarities\\ 
of the microcanonical entropy surface}
\author{D.H.E. Gross and E.Votyakov}
\address{
Hahn-Meitner-Institut
Berlin, Bereich Theoretische Physik,Glienickerstr.100\\ 14109 Berlin, Germany
and 
Freie Universit{\"a}t Berlin, Fachbereich Physik; \today}
\maketitle
\begin{abstract} 
It is discussed how phase transitions of first order (with phase
separation and surface tension), continuous transitions and
(multi)-critical points can be defined and classified for {\em
finite} systems from the topology of the energy surface $e^{S(E,N)}$ of the
mechanical N-body phase space or more precisely of the curvature
determinant $D(E,N)=\partial^2S/\partial E^2*\partial^2S/\partial
N^2-(\partial^2S/\partial E\partial N)^2$ {\em without taking the
thermodynamic limit}. The first calculation of the {\em entire
entropy surface $S(E,N)$} for a $q=3$-states Potts lattice gas on a
$50*50$ square lattice is shown. There are two lines, where $S(E,N)$
has a maximum curvature $\sim 0$. One is the border between the
regions in \{$E,N$\} with $D(E,N)>0$ and with $D(E,N)<0$, the other
line is critical starting as a valley in $D(E,N)$ running from the
continuous transition in the ordinary $q=3$-Potts model, converting
at $P_m$ into a flat ridge/plateau (maximum) deep inside the convex
intruder of $S(E,N)$ which characterizes the first order liquid--gas
transition.  The multi-critical point $P_m$ is their crossing.
\end{abstract}
\noindent Boltzmann's gravestone has the famous epigraph:
\cent{\fbox{\fbox{\vecbm{$S=k*lnW$}}}} which puts thermodynamics on the
ground of mechanics \cite{einstein02,einstein04}. It relates the
entropy $S$ to the volume $W(E,N,V)=\delta\epsilon* tr\delta(E-H_N)$
of the energy ($E$) surface of the N-body phase space at given volume
($V$), the microcanonical partition sum. Here $\delta\epsilon$ is a
suitable small energy constant, $H_N$ is the $N$-particle Hamiltonian,
and
\begin{equation}
tr\delta(E-H_N)=\int{\frac{d^{3N}p\;d^{3N}q}{(2\pi\hbar)^{3N}}
\delta(E-H_N)}.
\end{equation}
The set of points on this surface defines the microcanonical ensemble
({\em ME}).

Today conventional thermodynamics is based on the canonical
statistical mechanics as introduced by Gibbs\cite{gibbs02}.  In the
thermodynamic limit {\em ThL} ($N\!\!\mra\!\!\infty|_{N/V=const}$)
the canonical ensemble ({\em CE}) is equivalent to the fundamental
({\em ME}) {\em if the system is in a pure phase} and the {\em ThL}
exists.

The fundamental difference of microcanonical thermodynamics ({\em
MT}) to conventional thermodynamics is that no non-mechanical
quantities like temperature, heat, pressure have to be introduced
a priori.

The link between {\em ME} and {\em CE} is established by Laplace
transform. E.g. the usual grand canonical partition sum is the double
Laplace transform of {\em ME}:
\begin{equation}
Z(T,\mu,V)=\doublint_0^{\infty}{dE\;dN\;e^{-(E-\mu
N)/T}tr\delta(E-H_N)}\label{grandsum}.
\end{equation}
This excludes all inhomogeneous situations, especially phase
separations.  There the entropy is non-extensive and the {\em CE}
contains several Gibbs states at the same temperature. Consequently, the
statistical fluctuations do not disappear in the {\em CE} even in the
thermodynamic limit. This is the reason why Gibbs himself excluded
phase separations in chapter VII of\cite{gibbs02} in a footnote, page
75. At phase transitions the {\em ME} and the {\em CE} describe
different physical situations.  If one combines a small system of
water at the specific energy $\epsilon_1$ of boiling water with a
large heat bath at $100^0$C it will remain at $100^0$C but may 
convert into steam in $50\%$ of the cases.  I.e. the fundamental
assumption used e.g. by Einstein \cite{einstein02} that our ``system
changes only by infinitely little'' does not hold.

It is important to notice that Boltzmann's and also Einstein's
\cite{einstein02,einstein04} formulation allows for {\em defining}
the entropy by $S_{micro}:=ln[W(E,N,V)]$ (in the following we
use $S_{micro}$ for $S(E,N)$ if it is not clear) as a single valued,
non-singular, in the classical case differentiable, function of all
{\em ``extensive'', conserved} dynamical variables. No thermodynamic
limit must be invoked and the theory applies to non-extensive systems
as well. Of course this is achieved by avoiding Gibbs-states,
``equilibrium states'' or ``most random'' states \cite{ellis85}.  On
the other hand fluctuations become then important and must be
simulated by Monte Carlo methods.  The microcanonical ensemble is the
entire microcanonical N-body phase space without any exception. In
{\em MT} the entropy is {\em not} ``a measure of randomness''
\cite{ellis85}, it is simply the volume $e^{S(E,N,V)}$ of the energy
surface. The latter point is extremely important as it allows to
address even thermodynamically unstable systems like collapsing
gravitating systems (for a recent application of {\em MT} to
thermodynamical unstable, collapsing systems under high angular
momentum see\cite{laliena98}).  In so far it is the most fundamental
formulation of equilibrium statistics\cite{ehrenfest12,ehrenfest12a}.
From here the whole thermostatics may be deduced. {\em MT} describes
how $e^{S(E,N,V)}$ depends on the dynamically conserved energy,
number of particles etc.. Of course we must {\em assume} that the
system can be found in each phase-space cell of $e^S$ with the same
probability.

Following Lee and Yang \cite{lee52} phase transitions are indicated by
singularities in $Z(T,\mu,V)$. Singularities of $Z(T,\mu,V)$, however,
can occur in formula (\ref{grandsum}) in the thermodynamic limit only
($V\mra\infty|_{N/V=\varrho,E/N=\varepsilon}$). For finite volume
$Z(T,\mu,V)$ is a finite sum of exponentials and everywhere
analytical. Only at points where {\em $S(E,N)$ has a curvature $\geq
0$} will the integral eq.\ref{grandsum} diverge in the thermodynamic
limit.  In these points, the Laplace integral \ref{grandsum} does not
have a stable saddle point. Here van Hove's concavity condition
\cite{vanhove49} for the entropy $S(E,N,V)$ of a stable phase is
violated. Consequently we {\em define} phase transitions also for
finite systems topologically by the points of non-negative curvature
of the entropy surface $S(E,N,V)$ as a function of the mechanical,
conserved ``extensive'' quantities like energy, mass, angular momentum
etc..

Experimentally one identifies phase transitions of course not by the
singularity of $Z(T,\mu)$ but by the interfaces separating coexisting
phases, e.g. liquid and gas, i.e. by the {\em inhomogeneities}.  The
interfaces have three effects on the entropy:
\begin{enumerate}
\item There is an entropic gain by putting a part ($N_1$) of the
system from the majority phase (e.g. liquid) into the minority phase
(bubbles, e.g. gas), but this is connected with an energy-loss due to
the higher specific energy of the ``gas''-phase,
\item an entropic loss proportional to the interface area by the correlations
between the particles in the interface, leading to the convex intruder
in $S(E,N,V)$ and is the origin of surface tension \cite{gross150},
\item and an additional mixing entropy for distributing the $N_1$-particles
in various ways into bubbles.
\end{enumerate}
At a (multi-) critical point two (or more) phases become
indistinguishable and the interface entropy (surface tension)
disappears.

Microcanonical thermostatics was introduced in great detail for
atomic clusters and nuclei in \cite{gross153}. In two further papers
we showed how the surface tension can quantitatively be determined
from the microcanonical $S(E,N,V)$ for realistic systems like small
liquid metal drops \cite{gross158,gross157}. Needless to say that
only by our extension of the definition of phase transitions to
finite systems it make sense to ask the question: ``How many
particles are needed to establish a phase transition?'' The
results we are going to show here will again demonstrate that often $N$
does not need to be large in order to see realistic values (close to
the known bulk values) for the characteristic parameters.  Other
examples where shown in our earlier papers\cite{gross150,gross158,gross157}.

In the following we investigate the 3-state Potts lattice-gas model
on a 2-dim $L^2=50^2$ square lattice.  We will see how the total
microcanonical entropy surface $S(E,N)$ decovers even the most
sophisticated thermostatic features as first order phase
transitions, continuous phase transitions, critical and multicritical
points even for finite systems and non-extensive systems. This
demonstrates that the microcanonical statistics is in contrast to
Schr{\"o}dinger's claim \cite{schroedinger44} quite able to handle not
only gases but also phase transitions and critical phenomena.
Important details like the separation of phases and the origin of
surface tension can be treated. The singularity of the canonical
partition sum at a transition of first order can be traced back to
the loss of entropy due to the correlations between the surface atoms
at phase boundaries
\cite{gross150}.

\begin{figure}
\includegraphics*[bb = 0 0 225 225, angle=-0,
width=7cm,clip=true]{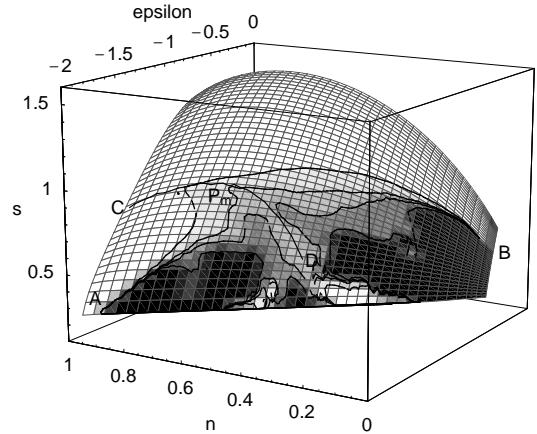}
\caption{
Entropy $s=\frac{1}{L^2}S_{micro}(\epsilon\!\!=\!\!E/L^2,n\!\!=\!\!N/L^2)$. The
grey levels are as in fig.\ref{det}: white regions: concave, pure
phases (in $A$$P_m$$C$ ordered, ``solid'', in $C$$P_m$$B$
disordered ``gas''); black: convex, phase separation
(``liquid--gas''); and light grey strips: critical branches; crossing:
multicritical point $P_m$. If one plots $S_{micro}$ vs. $\beta$ and
$\beta\mu$ like in {\em CT} the two wings $P_mA$ and $P_mB$ are
mapped onto one-another in the figure \ref{Sintens} below. The black
regions of the intruder get folded in between (see fig.\ref{sbtbms}).
\label{6d3_e}} 
\end{figure}
Briefly, a few words about our method which will be published in
detail in \cite{gross169}: The Hamiltonian is
$H=-\frac{1}{2}\sum_{i,j}\delta_{\sigma i,\sigma j}$ ($i,j$
nearest neighbors). We covered all space $\{E=\epsilon*L^2,N=n*L^2\}$
by a mesh with about $1000$ knots with distances of
$\Delta\epsilon=0.04$ and $\Delta n=0.02$. Due to our limited
computational resources (DEC-Alpha workstation) we could not use a
significantly denser mesh.  At each knot $\{\epsilon_i,n_k\}$ we
performed by microcanonical simulations ($\approx 2*10^8$ events) a
histogram for the probabilities $P(\epsilon_i,n_k)$ for the system to
be in the narrow region $(E_i\pm 4)*(N_k\pm 4)$ of phase space.  Local
derivatives $\beta=\left(\partial S(E,N)/\partial E\right)_N$,
$\beta\mu=-\left(\partial S(E,N)/\partial N\right)_E$ in each
histogram give the ``intensive'' quantities, so that the entire
surfaces of $S(E,N)$, $\beta(E,N)$, $\beta\mu(E,N)$ can be
interpolated. The first derivatives of the interpolated (smoothed)
$\beta(E,N)$ and $\beta\mu(E,N)$ give the curvatures.

The figure \ref{6d3_e} shows some of our recent results for
$S(E/L^2,N/L^2)$ for the case of the diluted $q=3$ Potts model. Grid
lines are in direction $[E-E_0(N)]/[E_{max}(N)-E_0(N)]=$const. resp.
$N/L^2=$const.. The black region is the intruder at the first-order
condensation transition (``liquid--gas coexistence'') with positive
largest curvature of $S(E,N)$.  This corresponds to the similar region
in the Ising lattice gas, respectively the original Ising model as
function of the magnetization. At the light grey strip $S(E,N)$ is
critical with vanishing largest curvature. The line from point $C$
over the {\em multicritical point} $P_m$ to $D$ corresponds from $C$
to $P_m$ to the familiar continuous transition in the ordinary $q=3$
Potts model. At $P_m$ this line crosses the rim of the intruder from
$A$ to $B$ which is the border of the first order transition. This
crossing determines the {\em multicritical point} $P_{m}$ quite well
at $\beta_m= 1.48\pm 0.03$, $\beta\mu_m= -2.67\pm 0.02$ or
$\epsilon_m\sim -1$, $n_m\sim 0.7$. From here the largest curvature
starts to become $\goo 0$. Naturally, $P_m$ spans a much broader
region in \{$\epsilon,n$\} than in
\{$\beta,\beta\mu$\}, remember here $S(E,N)$ is {\em flat}.
\begin{figure}
\includegraphics*[bb = 0 0 226 226, angle=-0, width=6cm,  
clip=true]{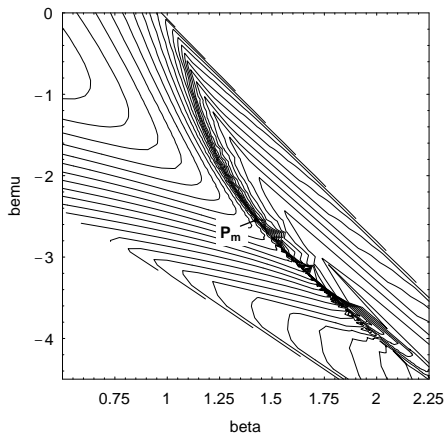}
\\~\\
\caption{
Contour plot of the entropy $s_{micro}(\beta,\beta\mu)$.  One clearly
sees the critical line where $s_{micro}(\beta,\beta\mu)$ becomes
tripelvalued and/or at the ordinary Potts continuous transitions
$\partial S/\partial\beta\sim -\infty$. Here several points of
$S(E,N)$ from the entropy surface are folded onto one-another. That
is the origin of the thick multi-valued (``critical'') line in
$s_{micro}(\beta,\beta\mu)$.\label{Sintens} }
\end{figure}

\begin{figure}
\includegraphics*[bb =0 0 225 225, angle=-0, width=6.5cm,  
clip=true]{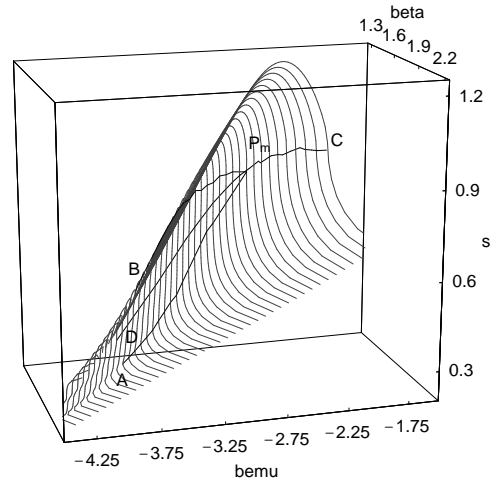}
\\~\\
\caption{$s_{micro }(\beta,\beta\mu)$, the same as fig.\ref{Sintens}.
The grey lines are for $\beta=$const.  The backbending of the intruder left
of $P_m$ which is hidden behind the thick critical line in
fig.\ref{Sintens} is clearly seen. This corresponds to the whole
intruder of phase separation (black and grey in fig.\ref{det}). The
lines connect the points of $\partial
s/\partial(\beta\mu)|_\beta=-\infty$ and thus correspond to the
critical lines in fig.\ref{det}. The positions of $A$,$B$,$D$,$C$ are
only roughly indicated.}
\label{sbtbms} 
\end{figure}

If one plots the entropy $s_{micro}(\beta,\beta\mu)$ as function of
the ``intensive'' variables $\beta\mu=-\partial S/\partial N$ and
$\beta=\partial S/\partial E$, we obtain picture \ref{Sintens}.  This
corresponds to the conventional grand-canonical representation if we
would have calculated the grand canonical entropy from the Laplace
transform  $Z(T,\mu,V)$, eq.\ref{grandsum}.  As there are several
points $E_i,N_k$ with identical $\beta,\beta\mu$,
$s_{micro}(\beta,\beta\mu)$ is a multivalued function of
$\beta,\beta\mu$. Here the entropy surface $S(E,N)$ is folded onto
itself see fig.\ref{sbtbms} and in fig.\ref{Sintens} these points
show up as a black critical line (dense region). The backfolded
branches of $S(E,N)$ are jumped over in eq. \ref{grandsum} and get
consequently lost in $Z(T,\mu)$. This demonstrates the far more
detailed insight one obtains into phase transitions and critical
phenomena by microcanonical thermostatics which is not accessible by
the canonical treatment.

\begin{figure}
\includegraphics*[bb =0 0 225 225, angle=-0, width=7cm,  
clip=true]{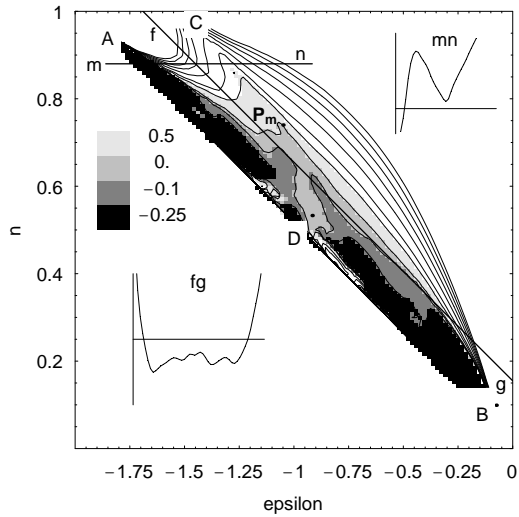}
\\~\\
\caption{Contour plot of the determinant of curvatures
$D(\epsilon\!\!=\!\!E/L^2,n\!\!=\!\!N/L^2)$.  In the two light gray
strips we have $D(\epsilon,n)\sim 0$. Here the transition is critical
(continuous). The crossing point $P_{m}$ indicates the multicritical
region. The two inlets show cuts along the line $f$\lra$g$
($\epsilon=-2n+0.28$) through the intruder with first order
transition and negative $D$ (positive largest curvature). To
understand the dramatic loss of information in the {\em CT}: The
whole black and grey region (about half of the entire phase space
!) gets lost, see also fig\ref{sbtbms}.  The other along $m$\lra$n$
near to the ordinary q=3 Potts model ($n=0.88$) shows to the left the
ordered phase with positive $D$, then the deep valley with nearly
vanishing curvature and to the right the disordered (``gas'') phase.
}\label{det} 
\end{figure}

In figure \ref{det} the determinant of curvatures of $S(E,N)$:
\begin{equation}
D(E,N)= \left\|\begin{array}{cc}
\frac{\partial^2 S}{\partial E^2}& \frac{\partial^2 S}{\partial N\partial E}\\
\frac{\partial^2 S}{\partial E\partial N}& \frac{\partial^2 S}{\partial N^2}
\end{array}\right\| \label{curvdet}
\end{equation}
is shown. On the diagonal we have the ground-state of the $2$-dim
Potts lattice-gas with $\epsilon=-2n$, the upper-right end is the
complete random configuration (not shown), with the maximum allowed
excitation $\epsilon_{rand}=-\frac{2n^2}{q}$. In the upper right
(white) $D>0$, both curvatures are negative. In this region the
Laplace integral eq.\ref{grandsum} has a stable saddle point. This
region corresponds to pure phases.

In the light gray region we have $D\sim 0$. This is the critical
region. Here the largest eigenvalue of $D$ is $0$. Two branches cross
here: One goes $\approx$ parallel to the ground state ($E\propto
-2N$) from $A$ to $B$. This is {\em a rim} in $D(E,N)$, the border
line between the region with $D(E,N)>0$, and the region with
$D(E,N)<0$ (black) where we have the first order liquid---gas
transition of the lattice-gas. The Laplace integral (\ref{grandsum})
has no stable saddle point and in the {\em ThL} the grand canonical
partition sum (\ref{grandsum}) diverges. Here we have a separation
into coexisting phases, e.g. liquid and gas. Due to the surface
tension or the negative surface entropy of the phase boundaries,
$S(E,N)$ has a {\em convex} intruder with positive largest curvature.

The other branch from $C$ to $P_m$ is {\em a valley} in $D(E,N)$.
Here the largest curvature of $S(E,N)$ has a local {\em minimum and
$D\sim 0$} (it would be $D=0$ with a higher precision of the
simulation), running from the point (near $C$) of the continuous
phase transition at $n=1$ and $\epsilon= -1.57$ of the ordinary
$q=3$-Potts model downwards to $P_m$. It converts below the crossing
point $P_m$ into {\em a flat ridge} inside the convex intruder of the
first order lattice-gas transition.  The area of the crossing of the
two critical branches $CP_mD$ and $AP_mB$ is the {\em multi-critical
region} $P_m$ of the $q=3$ Potts lattice gas model.

Conclusion: Microcanonical thermostatics ({\em MT}) describes how the
entropy $S(E,N)$ as defined entirely in mechanical terms by Boltzmann
depends on the conserved ``extensive'' mechanical variables: energy
$E$, particle number $N$, angular momentum $L$ etc.  This allows to
study phase transitions also in small and in non-extensive systems.  If
we define phase transitions in finite systems by the topological
properties of the determinant of curvatures $D(E,N)$
(eq.\ref{curvdet}) of the microcanonical entropy-surface $S(E,N)$: a
single stable phase by $D(E,N)>0$, a transition of first order with
phase separation and surface tension by $D(E,N)<0$ , a continuous
(``second order'') transition with $D(E,N)=0$, and a multi- critical
point where more than two phases become indistinguishable by the
branching of several lines with $D(E,N)=0$, then there are remarkable
similarities with the corresponding properties of the bulk
transitions.

The advantage of {\em MT} compared to {\em CT} is clearly demonstrated:
About half of the whole phase space, the intruder of $S(E,N)$ or the 
non-white region in fig.\ref{det}, gets lost in conventional canonical
thermodynamics. Without any doubts this contains the most sophisticated
physics of this system. 
Due to limited computer resources this could be demonstrated 
with only limited precision. We are convinced our conclusions will be
verified by more extensive -- and more expensive -- calculations.
\\~\\
\noindent Acknowledgment: D.H.E.G thanks M.E.Fisher for the
suggestion to study the Potts-3 model and to test how the
multicritical point is described microcanonically. We are
gratefull to the DFG for financial support.

\end{document}